\documentclass[fleqn,10pt]{wlscirep}

\usepackage[T1]{fontenc}
\usepackage{blindtext}
\usepackage{amsfonts}
\usepackage{amsmath}
\usepackage{microtype}
\usepackage{graphicx}
\usepackage{overpic}
\usepackage{booktabs}
\usepackage{tabularx}
\usepackage{algorithm}
\usepackage[numbers, compress]{natbib}
\usepackage[all]{hypcap}
\hypersetup{citecolor=blue}
\hypersetup{linkcolor=blue}
\hypersetup{urlcolor=blue}
\usepackage[nameinlink,capitalize]{cleveref}

\makeatletter
\newcommand{\settitle}{\@maketitle}
\makeatother

\usepackage{grffile}

\newcommand{\be}{\begin{equation}}
\newcommand{\ee}{\end{equation}}
\newcommand{\bea}{\begin{eqnarray}}
\newcommand{\eea}{\end{eqnarray}}

\Crefname{equation}{Eq.}{Eqs.}
\Crefname{figure}{Fig.}{Figs.}
\Crefname{tabular}{Tab.}{Tabs.}

\renewcommand{\ref}[1]{[\ref{#1}]}
\crefname{Methods}{Methods}{Methods}

\usepackage[caption=false]{subfig}

\graphicspath{{.}} 
\usepackage[normalem]{ulem} 

\makeatletter
\DeclareRobustCommand{\cev}[1]{%
  {\mathpalette\do@cev{#1}}%
}
\newcommand{\do@cev}[2]{%
  \vbox{\offinterlineskip
    \sbox\z@{$\m@th#1 x$}%
    \ialign{##\cr
      \hidewidth\reflectbox{$\m@th#1\vec{}\mkern4mu$}\hidewidth\cr
      \noalign{\kern-\ht\z@}
      $\m@th#1#2$\cr
    }%
  }%
}
\makeatother

\newcommand{\customsection}[1]{\phantomsection\def\@currentlabel{#1}\label{#1}\section{#1}}

\begin{document}

\title{Cohesive urban bicycle infrastructure design through optimal transport routing in multilayer networks}

\author[1,*]{Alessandro Lonardi}
\author[2,3,4]{Michael Szell}
\author[1]{Caterina De Bacco}
\affil[1]{Max Planck Institute for Intelligent Systems, Cyber Valley, T{\"u}bingen 72076, Germany}
\affil[2]{IT University of Copenhagen, 2300 Copenhagen, Denmark}
\affil[3]{ISI Foundation, 10126 Turin, Italy}
\affil[4]{Complexity Science Hub Vienna, 1080 Vienna, Austria}

\affil[*]{alessandro.lonardi@tuebingen.mpg.de}

\begin{abstract}
Bicycle infrastructure networks must meet the needs of cyclists to position cycling as a viable transportation choice in cities. In particular, protected infrastructure should be planned cohesively for the whole city and spacious enough to accommodate all cyclists safely and prevent cyclist congestion -- a common problem in cycling cities like Copenhagen. Here, we devise an adaptive method for optimal bicycle network design and for evaluating congestion criticalities on bicycle paths. The method goes beyond static network measures, using computationally efficient adaptation rules inspired by Optimal Transport on the dynamically updating multilayer network of roads and protected bicycle lanes. Street capacities and cyclist flows reciprocally control each other to optimally accommodate cyclists on streets with one control parameter that dictates the preference of bicycle infrastructure over roads. Applying our method to Copenhagen confirms that the city's bicycle network is generally well-developed. However, we are able to identify the network's bottlenecks, and we find, at a finer scale, disparities in network accessibility and criticalities between different neighborhoods. Our model and results are generalizable beyond this particular case study to serve as a scalable and versatile tool for aiding urban planners in designing cycling-friendly cities.\vspace{-0.5cm}
\end{abstract}

\maketitle

\section{Introduction}

Building cohesive urban networks of protected bicycle infrastructure is a contemporary challenge towards improving urban livability and transport, as cycling is a sustainable alternative to car-centric urban transport with massive economic and societal benefits \cite{buehler2016bikeway,gossling2019sca}. Most cities on the planet have negligible cycling infrastructure that needs to be built from scratch \cite{szell2022gub}. Nevertheless, certain cities do have a transport infrastructure that accommodates cycling well, which we can study to understand how close to an ``optimum'' they are in terms of serving the flow of cyclists or how much room there is for improvements. In particular, the Municipality of Copenhagen is known for its cohesive network of protected bicycle tracks and its corresponding cycling culture \cite{nielsen2013urban,haustein_comparison_2020,szell2022gub}, which has developed steadily over many decades \cite{carstensen2015spatio} and today provides well-quantified economic and health benefits to the city {and its population} \cite{fosgerau2023bikeability}.  As in Szell \emph{et al.} \cite{szell2022gub}, use the the Dutch CROW manual \cite{crow} to define what a  cohesive network is: A well-connected network that covers a large fraction of the city area. 

Despite these benefits, Copenhagen's bicycle network has mostly grown organically, and apart from aggregate projections \cite{ramboll_walking_2022} little is known about whether the existing infrastructure and its planned extensions \cite{kobenhavn2017,paulsen2023societally} fit the concrete local demands of cyclists. Furthermore, cyclist congestion has become a notorious phenomenon in the city because many of the permanent, physically separated bicycle tracks cannot accommodate all the cyclists using them. In fact, while the physical separation between raised bicycle tracks and roads provides safety to cyclists, this also makes the tracks difficult to widen \cite{vedel_bicyclists_2017,koubeck_breaking_nodate,haustein_comparison_2020}. In Copenhagen, high bicycle flows reportedly increase travel times. Since inadequate cycling lanes to accommodate traffic demands are a primary source of dissatisfaction for cyclists and fast travel times the main motivator for cycling, the Municipality of Copenhagen aims at reducing travel times by 15\% between 2016 and 2025 with policies such as ``overtaking lanes'' to make the cyclists' traffic flow smoother and electronic signals to update cyclists on congestion of roads \cite{city_of_copenhagen_technical_and_environmental_administration_bicycle_2016, city_of_copenhagen_technical_and_environmental_administration_bicycle_2019}.  While possibly less relevant in cities with fewer cyclists, this underscores the critical congestion issue in Copenhagen, where high cyclists flow can lead to over trafficking bicycle paths.

Identifying bottlenecks in the network, together with potential improvements, requires high-quality data \cite{ramboll_walking_2022,rahbekviero2023hgo}. In particular, cyclist flow data can be helpful in gauging the current levels of cyclist traffic in a city. However, flow data is often not collected systematically or at high enough resolution \cite{bhowmick2023systematic,lovelace2022jittering,fosgerau2023bikeability}. In practice, the availability of current flow data is also of little relevance, as infrastructure network planning frameworks tend to be static and do not take into account the co-evolution of flows and infrastructure \cite{gerike2022network,de_groot_design_2016}. Further, flow data can be interpreted as a reflection of currently available infrastructure, which does not paint an accurate picture of latent demand \cite{cervero2013bike,oecd_transport_2021}.

From a traffic flow perspective, when many cyclists are likely to use a specific street, such a street should be equipped with more cycling infrastructure. Conversely, streets with low expected cyclist demand require little to no additional infrastructure. Here, we account for these temporal dynamics of supply versus demand in terms of infrastructure, using an approach inspired by Optimal Transport theory \cite{villani2008optimal}, in line with recent data-driven approaches of bicycle network analysis to assess how ``bikeable'' a city's transportation network is \cite{nielsen2013urban,fosgerau2023bikeability,steinacker2022demand,reggiani2022understanding}. We apply our framework to the case study of Copenhagen. It allows us to ask: \emph{Given a certain preference for protected infrastructure over unprotected roads, where would cyclists travel?} And: \emph{What are the implications for the protected infrastructure? Specifically, where should Copenhagen widen it?}

Previously, such questions involving cyclist flow have been tackled from different perspectives, for example from demand modeling \cite{paulsen2023societally,mahfouz2023road}, empirical measurements \cite{skov-petersen_how_2018,fosgerau2023bikeability}, or from a network science approach where flows on the whole city are approximated via betweenness centrality \cite{szell2022gub,vybornova_automated_2023}. Here, instead, we combine the focus on the whole city with the treatment of bicycle paths as a \emph{dynamic} network that can respond to flows and that accounts for the travel priorities of cyclists. To do so, we first evaluate global metrics to assess bikeability. Specifically, we calculate two metrics for the whole network, the Detour (how much do I have to reroute my original path if I prefer to travel on bicycle tracks?) and the Overlap (how much can I stay on bicycle tracks to reach my destination?). Unsurprisingly, this evaluation shows good results since Copenhagen's bicycle network is well-developed. We then refine our investigation at the district level and find inequalities in the service provided by the city's neighborhoods, demonstrating that our flow-based method allows us to unveil results that are difficult to assess with static topological quantities only. Finally, we also identify infrastructure criticalities, i.e., bottlenecks where cyclist congestion can be expected. As our study accounts for the dynamics between flows and infrastructure, it thus complements research efforts where the network's infrastructure is studied only statically \cite{rahbek_viero_connectivity_2020,vybornova_automated_2023}.

\paragraph{Methods.} The method we develop to predict cyclist routes is based on Optimal Transport theory \cite{villani2008optimal}. This powerful mathematical framework formalizes the problem of finding the cheapest trajectories connecting two distributions, which in our setup are paths taken by cyclists entering and exiting the city's streets. In the last decade Optimal Transport garnered significant popularity thanks to its computational efficiency \cite{peyre2019computational,cuturisinkhorn} and its flexibility to model tasks across several fields of science, such as genomics \cite{demetci2020gromov}, computer vision \cite{werman1985distance,peleg1989unified,rubner1998metric,rubner2000earth}, or protein folding \cite{koehl2019optimal}, among others. In this work, we formulate a set of adaptation equations that connect cyclist travel paths with time-dependent capacities. The latter variables represent bicycle track widths that evolve in time to allocate cyclists optimally, i.e., along their cheapest routes. The connection between dynamical systems and Optimal Transport has been extensively studied in the literature \cite{facca2019numerics,facca2016towards,facca2020branching,facca2021fast,bonifaci2012physarum,bonifaci2013short,baptista2022convergence,baptista2020network,leite2022revealing}. Here, we leverage such results to formulate principled adaptation equations that, at convergence, return exactly the shortest paths of cyclists. There are several advantages of employing Optimal Transport-based dynamical systems. First, convergence is achieved fast \cite{facca2021fast,lonardi2023immiscible}, giving an edge over traditional schemes such as Dijkstra's algorithms \cite{dijkstra1959note} (see also Methods in Supplementary Information). Furthermore, such methods offer significant flexibility across various modeling scenarios. For example, dynamical systems have been employed in multicommodity problems where different types of users interact when moving along the network edges \cite{lonardi2021designing,lonardi2021multicommodity, bonifaci2021physarum,adinoyi2021optimal,adinoyi2022sustainable}, or in setups where entry and exit inflows vary in time, or stochastically \cite{lonardi2021infrastructure,hu2013adaptation,katifori2010damage,corson2010fluctuations}. Recent extensions of such models include bilevel optimization \cite{lonardi2023bilevel}, and integrate engineering constraints into the optimization setup \cite{ibrahim2023optimal}.

\paragraph{Results.} Putting the model into practice by simulating cyclist routes on Copenhagen's transportation network \cite{dataset}, we explore a series of promising results for improving the city's accessibility to cycling infrastructure. We systematically study a series of computational experiments where cyclists are encouraged to prioritize varying degrees of safety over trip convenience (the latter being measured as their total path length). To compare our findings from Copenhagen with less developed networks, we extract optimal cyclist routes not only on the whole Copenhagen network but also on synthetic underdeveloped bicycle networks subsampled from the full Copenhagen network. A further comparison is Zürich, a city of similar size and population but with a much less developed bicycle network (7\% of Copenhagen's length, see \Cref{tab: table networks params}) serving as an example of a typical European city that largely neglects bicycle infrastructure despite its benefits \cite{szell2022gub}.

Our experiments focus on Copenhagen's efforts to build a cohesive bicycle network that can accommodate the needs of cyclists both in terms of safety and efficiency. Specifically, cyclists should be able to cycle without having to take lengthy detours while also having access to protected bicycle infrastructure over roads that would otherwise be unsafe due to mixing with vehicular traffic. We observe, in line with previous studies \cite{nielsen2013urban,fosgerau2023bikeability}, that the Municipality of Copenhagen strongly benefits from its cohesive network of bicycle paths. Nevertheless, a fine-grained analysis of the city's accessibility to cycling infrastructure reveals potential concerns. Particularly, Copenhagen's districts exhibit substantial disparities in the quality of service provided to citizens. Established districts such as Bispebjerg, Indre By, Østerbro, and Brønshoj-Husum allow cyclists to travel safely (larger usage of protected bicycle lanes) and conveniently (short travel times), as opposed to currently developing or less established areas such as Amager or Nordhavn, or districts like Frederiksberg that historically lag behind in network development \cite{carstensen2015spatio}.

Furthermore, we observe that traffic bottlenecks, assumed to arise from high cyclists' flow, critically affect the bridges connecting the island of Amager with the rest of the city. Such a result sheds light on a critical issue of the city's infrastructure, which is often neglected in urban planning, especially in organic bicycle network growth \cite{szell2022gub}: Connecting two largely populated hubs with few links can drastically amplify congestion. A solution to alleviate such congestion can be installing more links between hubs, such as Copenhagen's bicycle bridge ``Cykelslangen'' between Vesterbro and Amager which typically carries 19,100 cyclists on a weekday \cite{city_of_copenhagen_technical_and_environmental_administration_bicycle_2019}.

\section{Methods}

\begin{figure}[htpb]
    \centering{
    \includegraphics[width=0.88\textwidth]{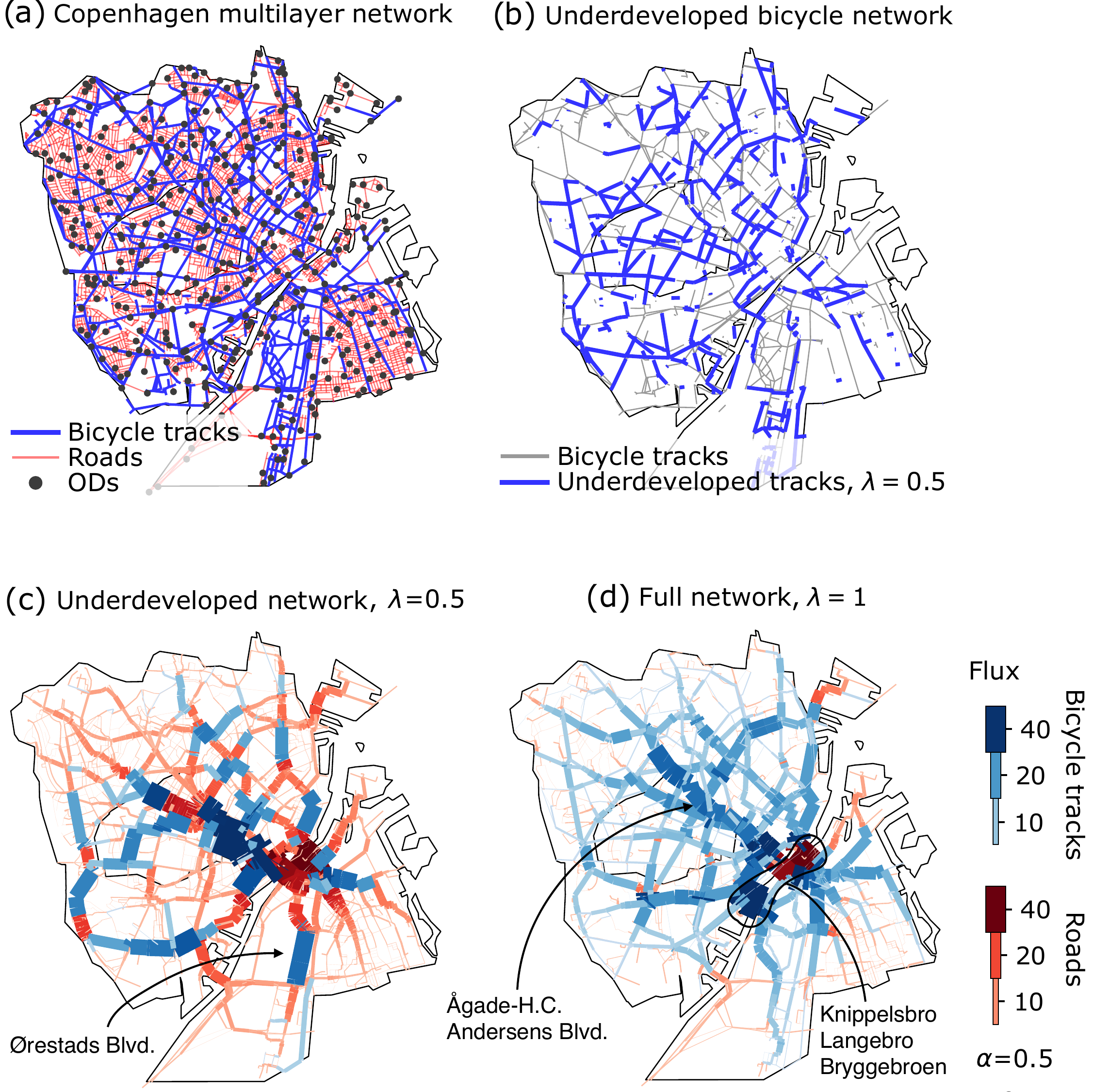}
    \caption{\textbf{Copenhagen's multilayer network of streets and protected bicycle tracks, the experimental setup, and simulation.}
    \textbf{(a)} Copenhagen's existing multilayer network of roads (red) and the full bicycle network (blue) with Origins-Destinations (ODs) distributed uniformly over the roads' junctions; Frederiksberg is included.
    \textbf{(b)} Synthetic underdeveloped networks are built by iteratively adding bicycle nodes randomly, together with their inter-layer and bicycle links, to the city's roads. In blue, we draw the bicycle tracks that are present in the displayed underdeveloped network; in gray, we draw edges that belong to the full city's network but not to the underdeveloped synthetic one.
    \textbf{(c)} Routing on the underdeveloped network ($\lambda=0.5$) with $\alpha=0.5$. The removal of 50\% of the bicycle node (approximatively corresponding to a reduction of 61\% of the tracks' length, see table \Cref{tab: table networks params}) leads to a strong concentration of cyclists on the road network. Fluxes $\sum_i |F_e^i|$ of cyclists that travel on roads (red) and bicycle tracks (blue) are visualized thicker and darker for more populated edges. Because of the choice of $\alpha<1$, cyclists tolerate slight detours in exchange for the safety of the protected tracks and thus populate the bicycle network preferentially.
    \textbf{(d)} Routing on the full network ($\lambda=1$) with $\alpha=0.5$. Since the bicycle network is entirely developed, fluxes are predominantly allocated on bicycle tracks instead of roads. Highlighted areas are those discussed in \Cref{ssec: global metrics}.}
    \label{fig:network topologies}
    }
\end{figure}

Before we introduce the mathematical formalism, we summarize our approach in words. Our starting point is the multilayer network of Copenhagen composed of 1) the road network and 2) the network of protected bicycle tracks. On this multilayer network, we simulate cyclist flows between uniformly distributed origins and destinations, \Cref{fig:network topologies}(a), by dynamically updating edge capacities following Optimal Transport-inspired adaptation rules.

The rationale behind and the functioning mechanism of such adaptation rules is that if many cyclists were expected to cycle on a street, then an adaptive city would build or widen bicycle tracks along that street. Conversely, if no cyclists were expected there, bicycle tracks of an adaptive city would be narrowed or removed. In our model, this feedback mechanism between cyclist demand and infrastructure adaptation is implemented automatically.  The cyclist flux, i.e., the number of cyclists moving between the network nodes, controls the infrastructure adaptation, i.e., the narrowing/widening bicycle tracks. This automatic update converges step by step to a weighted multilayer network of roads and bicycle tracks with an optimal design for accommodating cyclists, yielding capacities that allocate flows of cyclists along their most convenient paths.

The adaptation equations extract cyclist routes starting from their entry and exit inflows in nodes. Then, they allocate flows and capacities using one preference parameter $0\leq\alpha\leq1$ which controls the appeal of bicycle tracks versus roads, where $\alpha = 1$ means equal appeal and $\alpha < 1$ means more appeal for bicycle lanes. Following the literature, which reports detour factors between 1 and 2 for protected bicycle infrastructure \cite{reggiani2022understanding}, we systematically explore variations of $\alpha$ in the corresponding range $0.5 \leq \alpha \leq 1$. Although previous studies in Copenhagen have used proxy data for estimating empirical cyclist flows \cite{fosgerau2023bikeability}, here we assume no cyclist flow data is available as this is generally the case in most cities and makes our approach generally applicable.

\subsection{Optimal Transport dynamics for routing on networks}
\label{ssec: ot dynamics}

In order to find the Origin-Destinations (ODs) shortest paths on a multilayer network, we develop a set of adaptation rules controlling cyclist flows along edges to minimize their travel costs. We consider a multilayer network $G=(V_0 = \bigcup_k V_k, E_{0} = \{\bigcup_k E_{k}\} \cup \{\bigcup_{kk'} E_{kk'}\}$), where $V_k$ and $E_k$ are nodes and intra-layer edges of layers $k=1,\dots,L$, and $E_{kk'}$ are inter-layer edges between $k$ and $k'$. We suppose that each edge has length $\ell_e > 0$ and cost $w_e > 0$. In principle, $\ell$ and $w$ are not related; for example, in highways where passengers pay tolls $w$ to travel, prices of roads may be independent of their lengths and only determined by exogenous variables. However, for our case study, we suppose that $w$ depends on $\ell$ (we discuss our modeling assumption and its interpretation in \Cref{ssec: copenhagen network}).

Groups of cyclists $i = 1,\dots, M$ that enter and exit ODs are stored in a mass matrix $S$. Its entries are positive if $v=O^i$, i.e., the node $v$ is an entry node for the group $i$, and negative if $v \in D^i$, that is, if $v$ is one of the possible exit nodes for $i$. When $v$ is neither an entry nor an exit node for any group, the mass matrix equals zero. We also assume that the system is isolated, which translates to the condition $\sum_v S_v^i = 0$ for all $i$, namely that cyclists entering the network must also exit. The flow of cyclists of a group $i$ between two nodes $u,v$ is indicated by the flux $F_e^i$, with $e=(u,v)$.

The idea behind our model is that edges are endowed with adaptive capacities $\mu^i_e \geq 0$ that evolve in time to optimally allocate cyclists on their most convenient paths by means of a feedback mechanism. Such capacities admit a physical interpretation and can be thought of as road share used by cyclists of group $i$. This means that if a high flux $F_e^i$ of cyclists moves along edge $e$, its $\mu_e^i$ increases to meet travel demands, i.e., to enable them to move on $e$. Conversely, if there are no cyclists, the edge capacity decays to zero. By appropriately designing adaptation equations that connect fluxes and capacities, we are able to control $\mu$ in such a way that fluxes converge to the cyclists' weighted $w$-shortest paths. In addition, this formalism automatically accounts for a global constraint that bounds the capacities based on a total budget (for a detailed mathematical description of the adaptation equations, see Methods in Supplementary Information). 

One advantage of our method is the efficient calculation of the minimizer of $J(w) = \sum_{i,e} w_e |F_e^i|$, i.e.~the total weighted shortest path distance traveled by the cyclists \cite{facca2021fast,lonardi2023immiscible} (for details, see Methods in Supplementary Information). Moreover, dynamical systems can be generalized for modeling several complex routing tasks such as multicommodity problems \cite{lonardi2021designing,lonardi2021multicommodity, bonifaci2021physarum,adinoyi2021optimal,adinoyi2022sustainable}, setups with time-varying or stochastic loads \cite{lonardi2021infrastructure,hu2013adaptation,katifori2010damage}, and bilevel optimization \cite{lonardi2023bilevel}. They also enable the natural integration of infrastructure constraints, e.g., limiting the maximum road capacity, into the optimization setup \cite{ibrahim2023optimal}.

Other methods have been employed to extract cyclist paths and to model the optimal design of Copenhagen's bicycle tracks with their impact on the city's livability. Examples are reinforcement learning algorithms \cite{paulsen2023societally} to maximize a ``Net Present Value'' defined as a combination of objectives, including the infrastructure's construction and maintenance costs or the health benefits to the network's users. Additionally, nonlinear optimization routines \cite{yao2023perturbedutilitystochastictraffic,fosgerau2023bikeability,fosgerau2022perturbed} efficiently simulate realistic traffic flows on potential bicycle tracks that may be incorporated into the city's network in the future. Such models largely differ from ours based on adaptation equations. However, they offer an opportunity to enhance dynamical systems by integrating additional constraints into their formulation, thereby enabling the prediction of cyclist behaviors in a wide set of case studies. We do not address these nuances here.

\begin{table}[htpb]
\resizebox{\textwidth}{!}{%
\setlength{\extrarowheight}{.2em}
\centering{
\begin{tabular}{ c |  c  | c | c |  c  | c  } \toprule
 \textbf{Network} & \textbf{Length roads} & \textbf{Length bicycle} & \textbf{Coverage roads} & \textbf{Coverage bicycle} & $\boldsymbol{\Delta}$ \\
 & \textbf{[km]} & \textbf{tracks [km]} & \textbf{[\%]} & \textbf{tracks [\%]} & \textbf{[\%]} \\ \midrule
Copenhagen full network & 1103 & 420 & 72.4  & 27.6 & 100 \\
Copenhagen, $\lambda=0.9$ & 1103 & 311 $\pm$ 5 & 78.0 $\pm$ 0.3 & 22.0 $\pm$ 0.4 & 74 $\pm$ 5 \\
Copenhagen, $\lambda=0.75$ & 1103 & 254 $\pm$ 5 & 81.3 $\pm$ 0.3 & 18.7 $\pm$ 0.4 & 60 $\pm$ 5 \\
Copenhagen, $\lambda=0.5$ & 1103 & 162 $\pm$ 5 & 87.2 $\pm$ 0.4 & 12.8 $\pm$ 0.4 & 39 $\pm$ 5 \\
Zürich full network & 726 & 29 & 96.2 & 3.8 & 7 \\ \bottomrule
\end{tabular}
}
}
\caption{\label{tab: table networks params} \textbf{Parameters of the networks used in the numerical simulations.} The Coverage is the fraction between the length of the bicycle tracks (resp. of the roads) and the total length of all the network's streets. The parameter $\Delta$ is the fraction between the underdeveloped networks' bicycle tracks' length and the length of the tracks of the full network of Copenhagen. Mean values and standard deviations are computed for each $\lambda$, over 10 random seeds used to build the underdeveloped networks.}
\end{table}

\section{Results}

\subsection{Copenhagen's transportation network}
\label{ssec: copenhagen network}

{We consider the transportation network of Copenhagen with two layers extracted from \cite{dataset}}. The first contains $|E_{1}| = 11313$ roads that cover approximately $1100$ km ($72 \%$ of the total infrastructure), and the second is made of $|E_{2}| = 2554$ protected bicycle tracks that are $420$ km long (the remaining $28 \%$). The two infrastructures are joined by $|E_{12}| = 960$ inter-layer edges that connect nodes in layers $k=1$ and $k=2$ with the same coordinates. Such links are assigned negligible lengths and costs compared to any intra-layer edge. The network is made of $|V_0| = 9974$ nodes that represent junctions between streets, $|V_1| = 7825$ of them are road intersections (approximately $78 \%$ of the total), and $|V_2| = 2149$ are junctions between bicycle tracks (approximately $22 \%$).

Shortest path routing is performed between $M = 397$ groups of cyclists distributed uniformly in road junctions extracted from a rectangular tessellation of the city (see Methods in Supplementary Information). Each group $i$ enters one of the origin nodes, $O^i$, and exits uniformly from $M-1$ destination nodes (being the origin nodes of all other groups $j$, i.e., $D^i = \{ O^j \}^{j \neq i}$). The OD setup is modeled by a mass matrix with entries $S_v^i= +1$ for $v=O^i$, and $S_v^i= -1/(M-1)$ for all $v \in D^i$. Each entry represents the density of cyclists wanting to move from $O^i$ to all nodes of $D^i$. For a visualization of Copenhagen's multilayer network, see \Cref{fig:network topologies}(a).

As anticipated in \Cref{ssec: ot dynamics}, while links are assigned physical lengths $\ell$, we seek to optimize a different objective, the \mbox{$w$-weighted} total travel cost $J(w)$. This quantity corresponds to the geographical path length if $w=\ell$. However, its interpretation varies in accordance with the meaning of $w$. In our case study, we consider $w$ to represent the effective length of streets. For roads, we take $w = \ell$, whereas for bicycle tracks, we set $w = \alpha \,\ell$, with $0.5 \leq \alpha \leq 1$ being a dimensionless coefficient. This allows modeling scenarios where cyclists find traveling on bicycle tracks more appealing than on roads, ideally when safety is prioritized ($\alpha < 1$) over traveled length.

Beside simulating cyclist routing on Copenhagen's existing transportation network itself, we synthetically create a series of reduced networks that emulate less developed infrastructures than the existing one. We create these underdeveloped networks by iteratively adding randomly extracted bicycle nodes and all their links to a network with Copenhagen's roads and no bike lanes, while making sure that the network stays fully connected. Here, we use a control parameter $\lambda$ to interpolate between $\lambda = 0$, the network of roads without any bicycle infrastructure, and $\lambda = 1$, the network of roads and all bicycle infrastructure. Values of $\lambda$ in between correspond to an addition of a fraction $\lambda$ of randomly selected bicycle nodes. For each value of $\lambda = \{0.5, 0.75, 0.9 \}$, we generate $10$ networks. One such underdeveloped network for $\lambda = 0.5$ is visualized in \Cref{fig:network topologies}(b). In addition to the City of Copenhagen, we perform our simulation on the multilayer network of bicycle tracks and roads of Zürich, to show the decline in performance for a city where the bicycle network is strongly underdeveloped. It also serves as a validation for the proxy we chose to build underdeveloped synthetic networks. We list the network properties of all the different scenarios in \Cref{tab: table networks params}.

Fluxes $\sum_i |F_e^i|$ on one of Copenhagen's underdeveloped synthetic networks ($\lambda=0.5$) and on the city's full network ($\lambda=1$) are shown in \Cref{fig:network topologies}(c)-(d). The impact of the effective lengths on routing due to $\alpha < 1$ is visibly reflected by the larger concentration of cyclists on bicycle tracks. High fluxes occupy most of the cycling infrastructure, with scarcely populated roads. Furthermore, since on the underdeveloped network, bicycle tracks are fewer than those of the full network, congestion of both bicycle paths and roads is starker for $\lambda=0.5$. This happens because cyclists, with fewer bicycle tracks to travel on, concentrate on critical arteries of the cycling infrastructure while also cycling on roads when the network does not provide sufficient coverage (see Supplementary Information for additional network visualizations with different values of $\lambda$ and $\alpha$ and for Zürich). For the full network, \Cref{fig:network topologies}(d), the biggest traffic bottlenecks are the three main bridges Bryggebroen, Knippelsbro, and Langebro; the roads with the highest flux are in Ågade--H.C.~Andersens Blvd. In contrast, since Bryggebroen is pruned from the bicycle network of \Cref{fig:network topologies}(c), with $\lambda = 0.5$, the Ågade--H.C.~Andersens Blvd. shows increased flow, which now extends to Amager's Ørestads Blvd.

\subsection{Global metrics for bikeability}
\label{ssec: global metrics}

We develop and study a series of global metrics for fluxes with the goal of quantifying and shedding light on cyclist routing on Copenhagen's transport network.

\begin{figure}[htpb]
    \centering
    \includegraphics[width=1.0\textwidth]{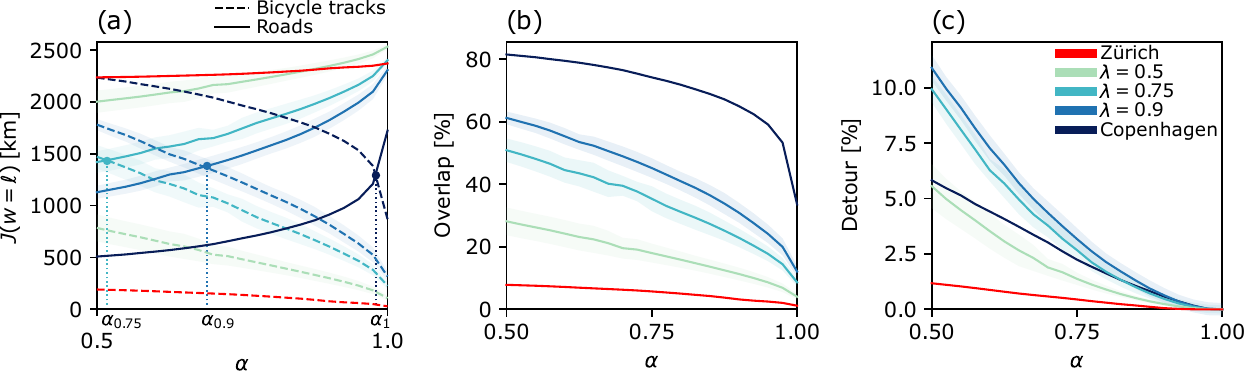}
    \caption{\textbf{Traveled distance, Overlap, Detour.} Results for $\lambda = 0.5, 0.75, 0.9$ are averaged over $10$ random network topologies, and shaded areas are their standard deviations. \textbf{(a)} $J(w=\ell)$ is solid for roads and dashed for bicycle lanes. Vertical dashed lines highlight points $\alpha_\lambda$ of crossing between two regimes where cyclists prefer to travel on bicycle tracks or on roads. \textbf{(b)}-\textbf{(c)} Overlap and Detour are computed over the whole multilayer network made of bicycle tracks and roads as defined in Methods in Supplementary Information.}
    \label{fig:global metrics}
\end{figure}

First, we observe the profiles of total traveled length $J(w=\ell)$ on bicycle tracks and roads for different values of $\alpha$ and $\lambda$, see \Cref{fig:global metrics}(a). The total traveled distance on roads decreases as $\alpha$ reduces (solid lines), and conversely the total traveled distance on bicycle tracks grows as $\alpha$ decreases (dashed lines). This trend is in accordance with the fact that a lower $\alpha$ corresponds to lower travel cost $w = \alpha \,\ell$ for bicycle tracks and, therefore, to cyclists willing to travel on longer cycle paths instead of shorter roads in order to maximize their safety (expressed by $J(w)$ with $\alpha <1$). The effect of higher $\lambda$ (green to blue) is to increase the traveled distance on bicycle tracks and to lower it on roads for all values of $\alpha$. This results from cyclists rerouting onto roads due to having fewer safe bicycle paths to travel on. In this respect, we observe that for $\lambda = 0.75, 0.9, 1$, bicycle lanes' and roads' traveled lengths cross at $\alpha = \alpha_\lambda$. This is a critical value separating two regimes where cyclists find it globally more convenient to cycle on bicycle tracks ($\alpha < \alpha_\lambda$) and on roads ($\alpha > \alpha_\lambda$). As $\lambda$ decreases, the crossing value $\alpha_\lambda$ gets lower. In fact, for underdeveloped bicycle networks, cyclists need to have higher incentives (lower $\alpha$) to favor bicycle tracks over roads. If the network is largely underdeveloped ($\lambda = 0.5$) and in Zürich, where bike lanes make only 3.8\% of the streets (see \Cref{tab: table networks params}), the traveled lengths do not cross at any $\alpha$ in the chosen range, showing that even with a large incentive for safety (a very low $\alpha$) cyclists have no other choice than to travel on roads. In a network where bicycle tracks have larger coverage than roads, i.e., where cycling on protected tracks is more convenient than traveling on roads without any incentive for safety ($\alpha = 1$), one should observe no crossing of the traveled lengths and the cost $J(w=\ell)$ for roads should always be below that of bicycle tracks.

In order to gain a deeper knowledge of cyclists' path trajectories, we calculate two additional metrics. The first is the Overlap, i.e., the percentage of the total length cycled on bicycle lanes. The second metric is the Detour, which is the relative change in traveled length between shortest-length fluxes (at $\alpha=1$) and the fluxes of cyclists who incentivize safety (at $\alpha<1$). For the mathematical definitions of Overlap and Detour, see Methods in Supplementary Information.

Results for Overlap and Detour are reported in \Cref{fig:global metrics}(b) and \Cref{fig:global metrics}(c), respectively. All Overlaps monotonically increase as $\alpha$ gets smaller, reflecting a progressively larger usage of the bicycle network. Since Copenhagen's bicycle tracks are well-developed, when $\alpha=0.5$, the Overlap surpasses $80 \%$, showing that most cyclists travel on bicycle tracks. The opposite happens in Zürich, where the Overlap is always below 10\%.

Detours have a less straightforward profile. Like the Overlap, they always increase as $\alpha$ decreases since cyclists reroute more when prioritizing safety. However, they are not monotone for $\lambda$. As one could expect, the Detour is higher for $\lambda = 0.9$ than for the full network of Copenhagen because cyclists have to reroute on longer paths when fewer protected tracks are available. However, for lower values of $\lambda = 0.75, 0.5$ and for Zürich, the Detour becomes progressively lower, falling below the full network for $\lambda = 0.5$ and Zürich. This decrease in Detour happens due to the bicycle infrastructure becoming \emph{too scarce}, where even high incentives to travel on bicycle tracks (low $\alpha$) are not sufficient for cyclists to travel on longer protected paths rather than on roads. See Supplementary Information for detailed visualizations of the cyclists' flows on all networks of Copenhagen and Zürich.

In summary, a city with a well-developed bicycle network should display both high Overlap and low Detour, allowing cyclists to safely ride their bicycles and arrive at their destinations conveniently. As demonstrated by Zürich, drawing conclusions using the Detour alone could be misleading since a low Detour with a low Overlap is symptomatic of a largely underdeveloped bicycle network.

\subsection{Bikeability of Copenhagen's districts}

In \Cref{ssec: global metrics}, we studied cycling patterns using global measures for the whole transportation network. Here, we aim to analyze the city at a finer scale, evaluating the contributions of individual districts to creating a safe and efficient bicycle network.

\begin{figure}[tbp]
        \centering{
        \includegraphics[width=1.0\textwidth]{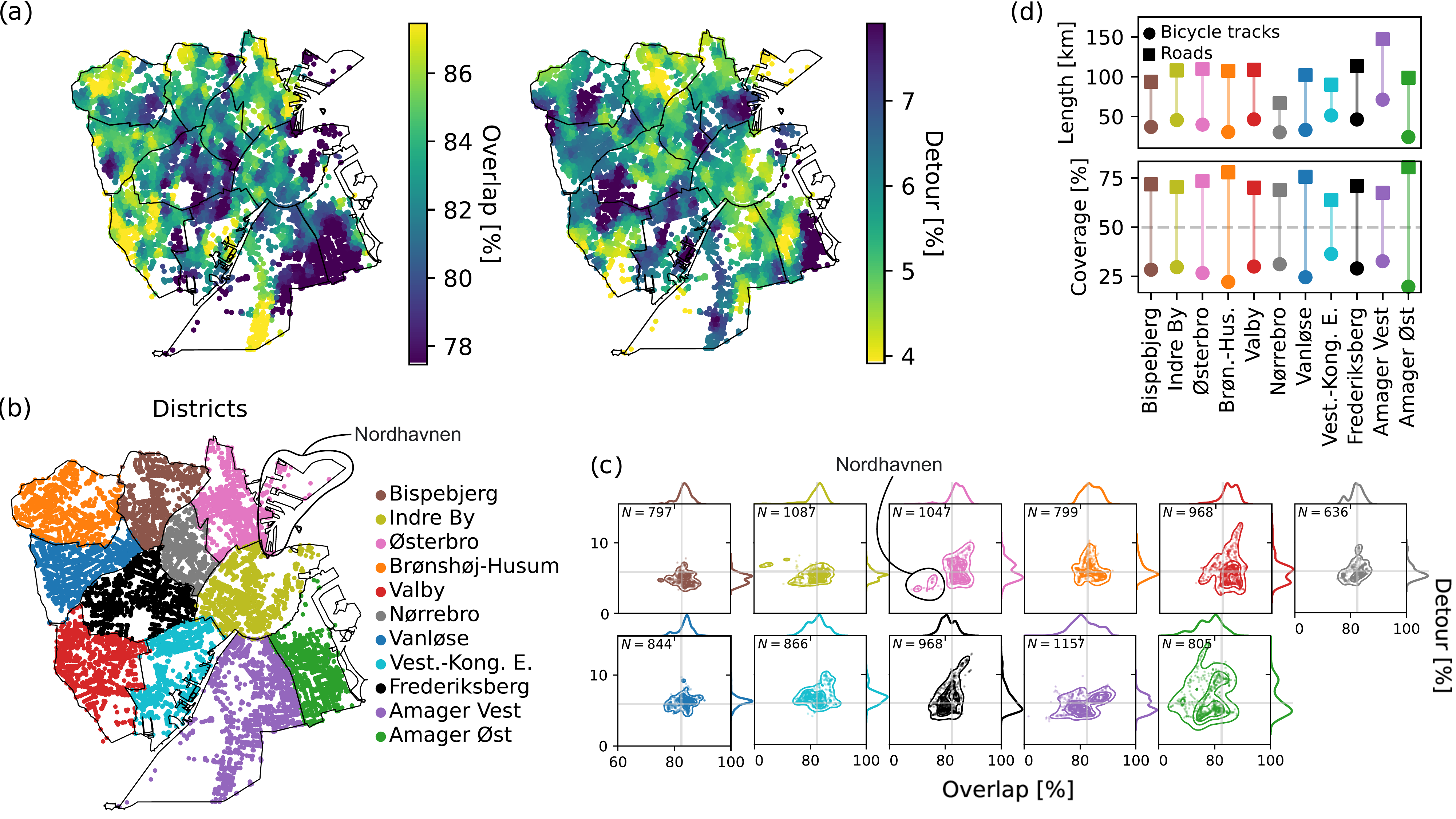}
        \caption{\textbf{Overlap and Detour for Copenhagen's districts.} \textbf{(a)} Distribution of Overlap and Detour on the nodes of the city. \textbf{(b)} Division of Copenhagen in districts. \textbf{(c)} Gaussian KDE plots for each district, colors correspond to the color scheme in (b), gray lines are averages of Overlap and Detour over the whole dataset. Contour lines are drawn to enclose $N = 200$ points at each level. (d) Absolute length in kilometers and Coverage of the infrastructure of each district, markers are colored as in (b).}
        \label{fig:overlap detour panel}
        }
    \end{figure}

We split the Municipality of Copenhagen into 11 parts, its 10 official districts \cite{copenhagen_districts}, and the Municipality of Frederiksberg, which is an independent enclave entirely surrounded by Copenhagen. In \cref{fig:overlap detour panel}(b), we show a map of the city with nodes colored according to the district they belong to.

First, we compute Detours and Overlaps for each street junction (for their mathematical definitions, see Methods in Supplementary Information). This allows us to get an estimate of which areas of the city provide the best service for cyclists, i.e., high Overlap and low Detour, or the worst, i.e., low Overlap and low Detour. Our analysis is performed with $\lambda=1$ and $\alpha=0.5$, thus simulating a scenario where cyclists are strongly encouraged to bike on cycling tracks over roads. Overall, the quality of service for cyclists in Copenhagen tends to be relatively high. In fact, on average, we get $\text{Overlap} \simeq 82 \%$ and $\text{Detour} \simeq 6 \%$ (see \Cref{tab: table overlap detours}). However, we find significant inequalities between districts.

Overlap and Detour distribute highly non-uniformly over the city's map, see \Cref{fig:overlap detour panel}(a), and vary within districts, see \Cref{tab: table overlap detours} and \Cref{fig:overlap detour panel}(c). The KDE plots (with a Gaussian kernel) in \Cref{fig:overlap detour panel}(c) show the distribution of Overlaps and Detours for the nodes of each district. Here, points in the bottom-right quadrant are those where the metrics are above the city's average values, i.e., those where cyclists can travel (more than on average) safely on protected bicycle lanes without having to bike on long paths. To quantify this idea of superior infrastructure, we label with $R$ the fraction between the bottom-right points and each district's total number of points, see \Cref{tab: table overlap detours}.

We first focus on currently developing or less established areas. For instance, in Amager (green and purple), a large portion of data points have small Overlaps and small Detours compared to the city's average ($R$ is below 9\%, i.e., worst overall performance). This low $R$ value signals that, under our model's assumptions, cyclists departing from Amager would rather travel on roads than protected tracks to avoid excessively long trips. Also, in Amager, cyclists leaving the southern area of the district use more bicycle tracks but have to take lengthy diversions. Detours and Overlaps in the geographically isolated port of Nordhavn, within Østerbro (pink), have a similar profile to Amager's, with cyclists traveling mostly on roads rerouting away from their shortest path.

More established areas, such as Bispebjerg (brown), Indre By (yellow), Østerbro (pink), and Brønshøj-Husum (orange) have high Overlaps. Remarkably, in Bispebjerg, all nodes are tightly concentrated around low Detour values, signifying a high and homogeneous quality of service in the district. This is also the district with the highest $R$ value. A notable exception is Frederiksberg. This district is one of the most affluent and well-established, so at first glance, it is surprising why it has a similarly low $R$ value ($R=8.88$) as Amager. However, Frederiksberg is a different municipality than Copenhagen, and therefore underlies a different policy-making process. It is known from historical records that Frederiksberg has been lagging behind Copenhagen with its bicycle network development \cite{carstensen2015spatio}, which could explain why it is also lagging behind in terms of Detour and Overlap.

These analyses are harder to quantify using infrastructural quantities only. We support this claim by plotting and listing the length of bicycle paths and roads for each district and their share in Coverage, i.e., how many streets correspond to one or the other type of infrastructure. These metrics are plotted in \Cref{fig:overlap detour panel}(d) and in \Cref{tab: table overlap detours} for decreasing $R$, and show no clear monotonicity trend. Examining the service of cycling infrastructure using only these infrastructural metrics would lead to potentially different conclusions than what was observed above using the estimated fluxes of passengers. For example, Amager Vest has better bicycle Coverage than Indre By, but the latter district has higher Overlaps and lower Detours. Thus, in principle, assessing which district is more bike-friendly is harder. For this specific example, one could argue that the reason that cyclists travel more on roads in Amager can be traced back to its extensive absolute length. However, a robust evaluation can only be made using metrics that consider various facets of the network's topology rather than a global metric only. We find that Detours and Overlap enable such an in-depth analysis at a fine scale.

\begin{table}[tpb]
\resizebox{\textwidth}{!}{%
\setlength{\extrarowheight}{.2em}
\centering{
\begin{tabular}{ c | c |  c  | c | c |  c  | c | c } \toprule
\textbf{District} & $\boldsymbol{R}$ & \textbf{Avg. Overlap} & \textbf{Avg. Detour} & \textbf{Length roads} & \textbf{Length bicycle} & \textbf{Coverage roads} & \textbf{Coverage bicycle}  \\ 
 & $(\downarrow)$ &  &  & \textbf{[km]} & \textbf{tracks [km]} & \textbf{[\%]} & \textbf{tracks [\%]} \\ \midrule
Bispebjerg & 69.76 & 83.39 & 4.85 & 93.45 & 36.86 & 71.71 & 28.29 \\
Indre By & 49.21 & 82.16 & 5.37 & 108.04 & 45.43 & 70.40 & 29.60 \\
Østerbro & 47.37 & 84.21 & 5.99 & 109.66 & 39.74 & 73.40 & 26.60 \\
Brønshøj-Husum & 44.81 & 83.82 & 6.03 & 107.18 & 30.38 & 77.92 & 22.08 \\
Valby & 44.31 & 84.86 & 5.88 & 108.59 & 46.32 & 70.10 & 29.90 \\
Nørrebro & 24.69 & 81.97 & 5.73 & 66.54 & 30.00 & 68.92 & 31.08 \\
Vanløse & 17.06 & 83.75 & 6.20 & 102.07 & 33.03 & 75.55 & 24.45 \\
Vest.-Kong. E. & 15.59 & 82.21 & 6.77 & 89.93 & 51.26 & 63.69 & 36.31 \\
Frederiksberg & 8.88 & 81.07 & 6.35 & 113.35 & 46.15 & 71.06 & 28.94 \\
Amager Vest & 8.47 & 81.34 & 5.63 & 146.96 & 71.11 & 67.39 & 32.61 \\
Amager Øst & 6.83 & 77.96 & 6.27 & 98.80 & 24.14 & 80.37 & 19.63 \\
Copenhagen & --- & 82.46 & 5.91 & 1102.62 (1144.59*) & 419.70 (454.43*) & 72.43 (71.58*) & 27.57 (28.42*) \\ \bottomrule
\end{tabular}
}
}
\caption{\label{tab: table overlap detours} \textbf{Overlap and Detour at districts' level and details on districts' infrastructure.} Districts are sorted with descending $R$, the fraction of points in a given district with both Overlap and Detour better than average. Districts' length of roads and bicycle lanes are computed by extracting all nodes of each district, taking the union of the edges to which they are connected, and computing their total length. This way, we tend to slightly overestimate the infrastructure length since we consider edges with only one link belonging to the chosen district. We denote with an asterisk the length of the Copenhagen Infrastructure computed by summing the length of all districts, and without an asterisk, that calculated directly from the city's network. Decimal digits are cut off to an arbitrary precision.}

\end{table}

\subsection{Traffic congestion}
\label{ssec: traffic congestion}

\begin{figure}[h]
        \centering
        \includegraphics[width=1\textwidth]{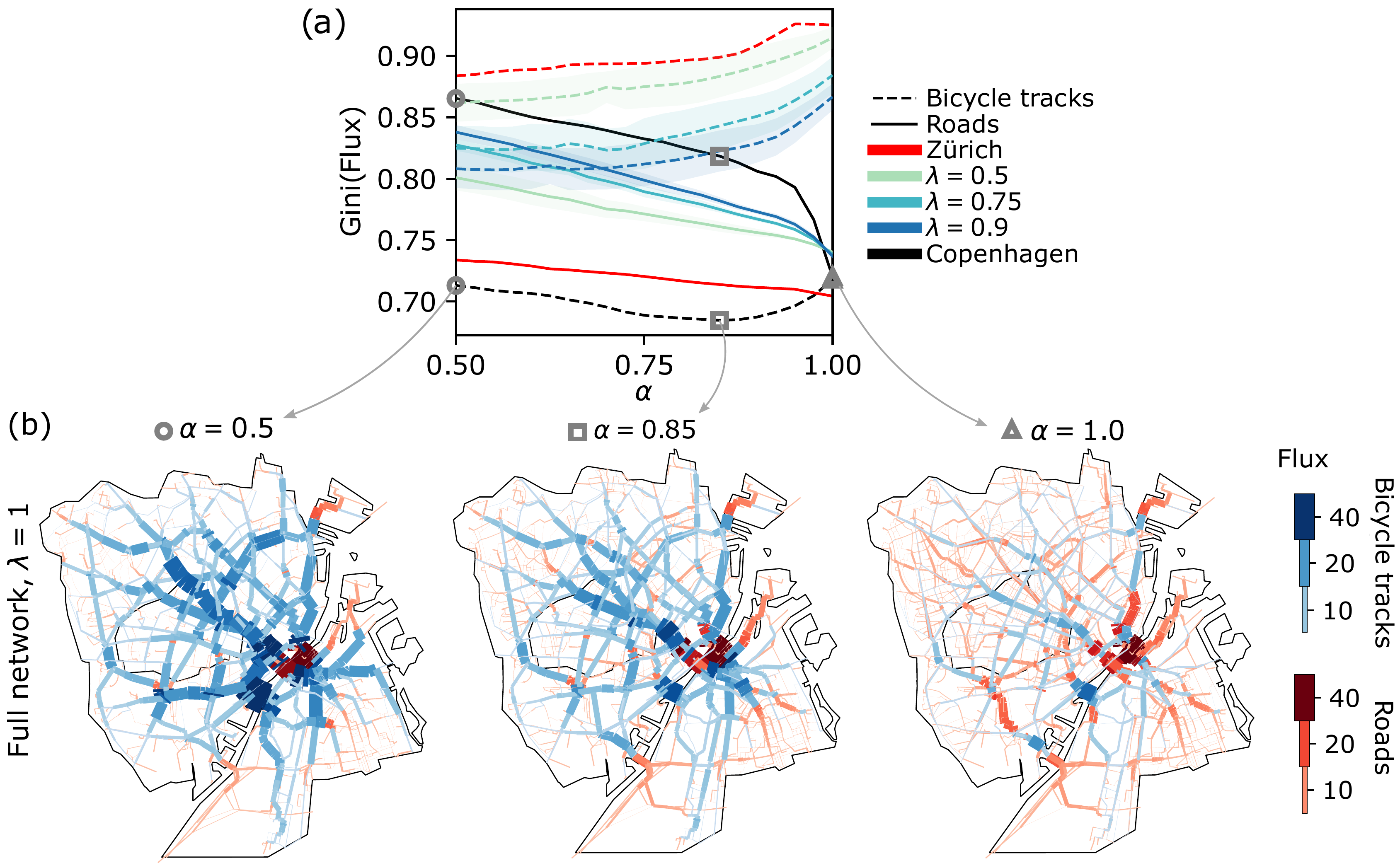}
        \caption{\textbf{Gini coefficient and transport networks at different congestion levels.} \textbf{(a)} The Gini coefficient is computed separately over the bicycle tracks and roads. Results for $\lambda = 0.5, 0.75, 0.9$ are averaged over $10$ random networks, and shaded areas are their standard deviations. \textbf{(b)} Networks with the flux of cyclists that travel on roads (red) and bicycle tracks (blue). Edges are thicker and darker when edges are more populated. Networks correspond to the three values of $\alpha$ denoted by markers (circle, square, triangle) in (a) and are shown for the full Copenhagen network ($\lambda=1$).}
        \label{fig:gini panel}
    \end{figure}

To conclude our analysis of Copenhagen's transportation network, we study cyclist congestion on its streets. Particularly, we evaluate the Gini coefficient of fluxes distribution ($\sum_i |F_e^i|$) on bicycle tracks and roads (for its mathematical definition, see Methods in Supplementary Information). A lower Gini coefficient signals uniformly distributed fluxes over the edges, whereas a high Gini coefficient is symptomatic of an uneven distribution. The coefficient is normalized so that $\text{Gini} = 0$ corresponds to uniform fluxes, and $\text{Gini} = 1$ to maximum inequality, i.e., all flux is concentrated on one edge. In the context of congestion, a low Gini index is preferable. We evaluate its profile against $\alpha$,  for all $\lambda$ values and for Zürich in \Cref{fig:gini panel}(a).

The Gini coefficient of the road and bicycle networks drastically differ. The former has a monotonically increasing profile as $\alpha$ decreases. This signals that traffic on roads becomes unbalanced when cyclists progressively reroute onto bicycle tracks. Bicycle tracks show different trends. For all synthetic underdeveloped networks with $\lambda = 0.5,0.75,0.9$ and for Zürich, we get an opposite pattern to those of the roads, i.e., the Gini coefficient monotonically \textit{decreases} with decreasing $\alpha$. However, the behavior is not monotonic in the full Copenhagen network. The Gini initially drops from its maximum value at $\alpha=0.5$ until a minimum at $\alpha=0.85$; then the Gini increases until reaching again its maximum value at $\alpha=1.0$, as shown in \Cref{fig:gini panel}(a).

This valley suggests that in a well-developed bicycle network at $\alpha = 1$, as cyclists are indifferent to choosing bicycle tracks over roads, bicycle tracks are largely unused by cyclists. This seeming preference could be explained by the phenomenon of \emph{parallel paths} where routing on bicycle tracks that go in parallel with a road is often slightly longer than a corresponding route on the road itself \cite{vybornova_automated_2023}. Subsequently, with lower $\alpha$, bicycle tracks get more uniformly populated as cyclists are encouraged to travel on them.  This holds until a critical value ($\alpha = 0.85$ for Copenhagen), after which an increasing number of cyclists travel on bicycle tracks, also causing more traffic congestion.

In \Cref{fig:gini panel}(b), we show how the bicycle network switches from being only lightly used to congested as we decrease $\alpha$ (right to left).
This congestion negatively impacts the quality and usability of the transportation infrastructure. In particular, we notice how the bridges connecting the larger portion of the city and the Amager island are the main traffic bottlenecks.

Finally, in \Cref{fig:gini panel}(a), we observe how congestion is further enhanced by the development status of the network. Underdeveloped infrastructures (lower $\lambda$) display significantly higher Gini than what is observed in the full Copenhagen network.

\section{Discussion}

While our analysis of Copenhagen's bicycle infrastructure revealed chokepoints and non-trivial local variations, it was a first step in applying Optimal Transport to bicycle network planning with potential open refinements. It also focused only on a short-term, flow-based approach that needs to be incorporated into a wider perspective of livable, sustainable, and just urban transport planning.

On a technical level, our model does not consider modal shares between bicycles and cars or between bicycles and other modes of transport. It could be that instead of rerouting onto longer paths, cyclists decide to start traveling via car -- a hypothesis that is not far-fetched, given recent developments of rising car use and stagnation of bicycle modal share in Copenhagen \cite{koglin2021cycling}. An option to model this scenario could consider multicommodity adaptation rules \cite{lonardi2021designing,lonardi2021multicommodity,ibrahim2023optimal}, or statistical physics approaches, such as Belief Propagation \cite{yeung2013physics,yeung2013networking,xu2021scalable}. Another open question is the effect of enriching the route choice model with more data: Further auxiliary data sets in this context are objective or subjective infrastructure quality and safety \cite{gossling2022subjectively,rahbekviero2023hgo} or personal routing preferences \cite{skov-petersen_how_2018,reggiani2022understanding,lukawska2023joint}. Further, we do not compare our simulated traffic flows with empirical data collected via tracking devices \cite{fosgerau2023bikeability,skov-petersen_how_2018}. While empirical flow data could make the discussion even more insightful, it comes with the potential bias of reflecting currently available infrastructure more than actual latent demand \cite{cervero2013bike,oecd_transport_2021}. One limitation of our uniform OD-matrix assumption could be surpassed via non-uniform mass inflows as in Refs.~\cite{steinacker2022demand,lonardi2021multicommodity,ibrahim2023optimal,morris2012transport} where inflows are extended, for instance, by assuming profiles based on node centralities. An alternative to build the OD matrix could be to add a distance decay term to the node inflows, which reduces the amount of cyclists traveling between distant ODs \cite{lovelace2017propensity}. Additionally, we note that while Overlap and Detour enable us to draw several quantitative insights about Copenhagen's cyclist flows and infrastructure, there are several other metrics that can be adopted, each potentially offering valuable insights \cite{tufail2016bicycle, arenalla2020developing}. 

In this work, we focused on simulating underdeveloped networks to assess the impact of less developed infrastructures than Copenhagen's on bikeability. However, a similar analysis could be performed by considering alternative scenarios, for instance by modeling future developments where additional bicycle tracks are included to improve bikeability. This would require thinking about how to strategically add additional network edges with growing strategies \cite{szell2022gub}. Finally, our analysis focuses on the optimal cyclists' fluxes under stationary entry/exit inflows. One could potentially include time-dependent demands that would allow to differentiate between short-term peak congestion and consistent bottlenecks. This could enable drawing a broader picture of traffic congestion on Copenhagen's streets, at the cost of higher model complexity \cite{lonardi2021infrastructure}.

Apart from advancing the field of flow modeling, on a higher level our study can pinpoint potential relations between bicycle infrastructure and socio-economic or policy factors. For a statistically reasonable regression analysis of our flow-based metrics with socioeconomic factors like income, education, or level of urban development, we unfortunately do not have enough spatial units available ($N=11$ districts). Nevertheless, we have reason to believe that the low $R$ values for Amager and Frederiksberg are not a coincidence. If the low $R$ value for Frederiksberg is indeed a remnant of the municipality's historical neglect of its bicycle network \cite{carstensen2015spatio}, it could provide a valuable lesson to the cumulative effects of policy-making. This potential effect remains open to be tested rigorously, as does the low performance in Amager and Nordhavn. It is yet unclear to which extent this performance is due to these districts' development status or their geographical remoteness. For example, should the city focus more on new connecting bicycle bridges like ``Cykelslangen'' or on extending Amager's own bicycle network to improve its situation?

Besides the theoretical research, our study has potential applications in urban planning. However, a notable caveat regarding technical feasibility needs to be discussed. In Copenhagen, it might not be realistic to dynamically widen protected bicycle tracks because the city is locked into a model of raised bicycle tracks \cite{vedel_bicyclists_2017,koubeck_breaking_nodate,haustein_comparison_2020,nieuwenhuijsen_implementing_2019}. This track infrastructure is safe, but it does not allow the flexibility of easy widening as in e.g.,~bollard-based approaches \cite{national_association_of_city_transportation_officials_urban_2014}. Therefore, the feasibility of the approach can depend on local contexts like the microscopic design principle of bicycle tracks. Nevertheless, our research highlights successfully the potential chokepoints in the system.

Additionally, we emphasize that while our model assumes cyclists choose OT-optimal paths, route preferences can be more nuanced and influenced by network, individual, and exogenous factors \cite{lukawska2024quantitative}. Integrating data-driven behavioral studies \cite{chou2023analysis, lukawska2024quantitative} into network models such as the one we considered here could provide valuable insights and help translate scientific findings into actionable strategies for making cities more bikeable. A possible step in this direction is the use of bilevel approaches \cite{lonardi2023bilevel}, where travelers' individual trajectories are taken into account by network managers.

In a wider perspective beyond a short-term flow-based study, long-term dynamics must be considered. Most importantly, it cannot be assumed that the total flow in the system is constant or independent of existing infrastructure. In the most extreme case of no existing protected bicycle infrastructure, cyclist flow can be assumed near zero because most people do not feel safe cycling \cite{hull2014bicycle}, as is the case in most cities worldwide \cite{szell2022gub,prieto2024abc}. Going with the ``build it and they will come'' principle \cite{cervero2013bike}, an increase in infrastructure then would lead to more flow in the system. Therefore, it is important to account not only for existing cyclists, because their numbers are strongly influenced by the existing (lack of) infrastructure. Instead, a network design is necessary that accounts for such future flow changes related to induced demand \cite{oecd_transport_2021}.

Ultimately, bicycle network analysis should also ask whether and why a bicycle network is even needed, or whether the goal of more livable, sustainable, and just cities \cite{gossling_urban_2016} can also be achieved via simpler means such as an overall reduction of speed limits and car parking, reallocation of street space from cars to sustainable transport, or introduction of congestion charges -- all particularly relevant for Copenhagen \cite{koglin2021cycling}. These interventions also come with the benefit of reducing the many dimensions of car harm \cite{miner_car_2024}. However, as such measures impose restrictions on motorized traffic and thus can imply a perceived unsettling change in the status quo and existing privileges, they depend on political choice. For cases where political leadership is not capable or willing to implement ``big bang'' interventions (sudden sweeping changes) \cite{nieuwenhuijsen_implementing_2019}, our research can help to fine-tune existing bicycle networks effectively via incremental interventions, and plan new networks or their extensions to account for cyclist congestion, to facilitate positioning cycling as a viable transportation choice in cities.

\section*{Data availability}
All data used for the experiments are publicly available \cite{dataset}.

\section*{Code availability}
An open-source Python code implementation is available \cite{gitrepo}.
\section*{Author contributions}
All authors contributed to conceiving the model and the experiments. A.L. conducted the experiments and produced all data visualization. All authors analyzed the results and reviewed the manuscript. All authors have read and agreed to the final version of the manuscript.

\section*{Competing financial interests}

The authors declare no competing interests.

\section*{Acknowledgments}

The authors thank the International Max Planck Research School for Intelligent Systems (IMPRS-IS) for supporting A.L. M.S.~acknowledges funding from EU Horizon Project JUST STREETS (Grant ID: 101104240) and from The Danish Ministry of Transport (Grant ID: CP21-033). All map data from OpenStreetMap.

\bibliography{bibliography}

\mbox{}
\clearpage
\newpage

\setcounter{equation}{0}
\setcounter{figure}{0}
\setcounter{section}{0}
\setcounter{table}{0}
\makeatletter
\renewcommand{\theequation}{S\arabic{equation}}
\renewcommand{\thefigure}{S\arabic{figure}}
\renewcommand{\thetable}{S\arabic{table}}
\renewcommand{\thesection}{S\arabic{section}}

\section*{\LARGE{Cohesive urban bicycle infrastructure design through optimal transport routing in multilayer networks: Supplementary Information}}

\section{Methods}

\subsection{Experimental setup}

\paragraph{Extraction of the ODs.}

Origins and destinations pairs (OD) are extracted as follows. We perform a rectangular tessellation with $25 \times 25$ tiles of the city of Copenhagen. We extract a random node from the road network for each tile containing at least a road junction. This procedure returns a total of $M = 397$ nodes. The extraction of nodes is constrained to those in the road network so that we can easily deconstruct the bicycle tracks and perform our experiments with underdeveloped tracks ($\lambda < 1$) while keeping the OD nodes fixed. A similar tessellation of size $27 \times 27$ is performed over the map of Zürich. This returns $M = 388$ OD nodes, to which we add $9$ random nodes to match the $M=397$ origin nodes used for Copenhagen. By assigning unitary inflowing mass to each origin node, we ensure that the number of people traveling in Copenhagen and in Zürich are the same.

\paragraph{Construction of synthetic bicycle networks.}

As mentioned in Section 2.2 (main text), the underdeveloped networks of Copenhagen are built by iteratively adding random bicycle nodes to the city's roads. In detail, for a fixed value of $\lambda$, we extract random nodes from the bicycle network, and we add them to the road network together with their inter-layer edges and bicycle edges, provided that the addition of the newly extracted node and edges return a fully connected network. We keep extracting nodes until the fraction of bike nodes of the underdeveloped networks surpasses a fraction $\lambda$ of the bicycle nodes of the full network of the city ($\lambda = 1$). A visualization of the underdeveloped bicycle networks for different values of $\lambda$ is in the Supplementary Information. For each value of $\lambda$, we perform experiments by ranging $\alpha$ between $[0.5,1.0]$, using a spacing of $w=0.025$.

\subsection{Optimal Transport adaptation equations}

\paragraph{Construction of the adaptation equations.}
Referring to the problem setup introduced in Section 2.1 (main text), we formalize the adaptation equations used to find cyclists' $w$-shortest paths on multilayer networks. Recall that the union of intra-layer and inter-layer edges is $E_{0} = \{\bigcup_k E_{k}\} \cup \{\bigcup_{kk'} E_{kk'}\}$, and the union of the nodes in all layers is $V_{0} = \bigcup_k V_k$. We solve the dynamical system
\begin{alignat}{2}
\label{eq: dynamics}
\frac{d \mu_e^i}{d t} &= \frac{(F_e^i)^2}{\mu_e^i}  - \mu_e^i \quad &&\forall e \in E_{0}, \forall i = 1,\dots,M \\
\label{eq: kirchhoff}
\sum_e B_{ve} F_e^i &= S_v^i \quad &&\forall v \in V_0, \forall i = 1,\dots,M \\
\label{eq: poiseuille}
F_e^i &= \frac{\mu_e^i}{w_e} (p_u^i - p_v^i) \quad &&\forall e \in E_{0}, \forall i = 1,\dots,M \,.
\end{alignat}
In \Cref{eq: kirchhoff} we write Kirchhoff's law, expressing conservation of mass. Here, we introduce a conventionally oriented network incidence matrix with entries $B_{ve} = +1$ if $v$ is the head of $e$, $B_{ve} = -1$ if it is its tail, and $B_{ve} = 0$ if it is neither of them. Making the incidence matrix terms explicit, at each node $v$ and for each $i$,  Kirchhoff's law can be expanded as $\sum_{e \in E^{\vec{v}}} F_e^i - \sum_{e \in E^{\cev{v}}}  F_e^i = S_v^i$. Here, we denoted with $E^{\vec{v}}$ the set of edges with $v$ as tail, and $E^{\cev{v}}$ those with $v$ as head.  Such a rewriting shows how local conservation of mass acts at each node, i.e., the difference between (positive) fluxes entering $v$ and (negative) exiting ones is equal to the net mass $S_v^i$. Since generally $|E_0| > |V_0|$, Kirchhoff's law in \Cref{eq: kirchhoff} is undetermined, we therefore assume that fluxes are differences of an auxiliary pressure potential $p_v^i$ as in \Cref{eq: poiseuille} (often referred as Hagen-Poiseuille equation in fluid dynamics, or Ohm's law in electromagnetism). In analogy with resistor networks, one could think of $F_e^i$ as the current intensity, of $p_v^i$ as the electric voltage, and of the ratio $\mu_e^i / w_e$ as the electrical conductance. Substituting \Cref{eq: poiseuille} into \Cref{eq: kirchhoff} allows us to solve Kirchhoff's law. In practice,  \Cref{eq: kirchhoff} goes from being overdetermined ($E_0$ unknown fluxes and $V_0$ input $S_v^i$ value for each $i$), to be a square system of linear equations $\sum_u L^i_{vu} p_u^i = S_v^i$ for each $i$. Here, $L^i$ is the $\mu^i/w$-weighted graph Laplacian. This allows us to obtain pressure potentials as $p_v^i = \sum_u (L^i)^\dagger_{vu} S_u^i$, where and $\dagger$ denotes the Moore-Penrose inverse.

By means of these substitutions, the feedback mechanism in \Cref{eq: dynamics} becomes dependent only on $\mu$.  Here, each $\mu^i_e \geq 0$ evolves in time to allocate cyclists with a feedback mechanism.  Precisely,  a high flux $F_e^i$ of on edge $e$, makes $\mu_e^i$ grow to allocate demand. Conversely, if there are no cyclists, i.e., $F_e^i=0$,  the capacity $\mu_e^i$ decays exponentially as \Cref{eq: dynamics} becomes ${d \mu_e^i} / {d t} = - \mu_e^i$. Solutions of  \Crefrange{eq: dynamics}{eq: poiseuille} can be found with a first-order numerical integrator, e.g., the forward Euler method, used in alternation with a sparse linear solver for Kirchhoff's law. At convergence, we get capacities $\mu^\star = \lim_{t \to + \infty} \mu (t)$ and asymptotic fluxes $F^\star = F(\mu^\star, p(\mu^\star))$.

\paragraph{Connection with Optimal Transport.} The crucial result of our method is that asymptotic fluxes $F^\star$ correspond to $w$-shortest paths $\pi^{\mathrm{sp}}$ between ODs \cite{bonifaci2012physarum, bonifaci2013short,lonardi2023immiscible}. Such a connection can be drawn by employing Optimal Transport theory. Precisely, in Optimal Transport the goal is to find the shortest path between $g^i$ and $h^i$, the cyclists' entry and exit distributions supported on the nodes, by solving the linear program (primal Kantorovich problem) \cite{kantorovich1960mathematical}
\begin{equation}
    \label{eq: ot problem}
    \min_{\Pi(g^i, h^i)} \sum_{u,v} w_{uv} \pi^i_{uv} \quad \forall i = 1,\dots,M\,.
\end{equation}
Here $\Pi(g^i, h^i)$ is the set of admissible transport paths $\pi^i$ that satisfy mass conservation $\sum_u \pi^i_{uv} = h^i_v$ and $\sum_v \pi^i_{uv} = g^i_u$ between entry and exit distributions of cyclists of group $i$. Transport path entries $\pi^i_{uv}
$ express the probability of routing cyclists of group $i$ on the a path between $u$ and $v$, with optimal cost $w_{uv}$, while satisfying the mass conservation constraints.

It can be proved \cite{facca2019numerics,lonardi2023immiscible} that \Crefrange{eq: dynamics}{eq: poiseuille} admit a Lyapunov functional $J^i(t) = \sum_{e} w_e |F_e^i[\mu(t), p(\mu(t)) ]|$ that converges to the minimum of \Cref{eq: ot problem}. Moreover, the search space defined by $\Pi(g^i, h^i)$ is exactly the set of fluxes satisfying Kirchhoff's law for each $i$. Therefore, by solving \Crefrange{eq: dynamics}{eq: poiseuille} we find asymptotic fluxes $F^\star$ that correspond to $w$-shortest paths $\pi^{\mathrm{sp}}$.

Similarly, the adaptation equations \Crefrange{eq: dynamics}{eq: poiseuille} also solve a constrained energy minimization problem with objective $\sum_{i,e} w_e |F_e^i|^2$ and whose constraints are Kirchhoff's law for each $i$, and $\sum_{i,e} w_e \mu_e^i$, fixing the total budget to build the infrastructure, or ``volume'' of the network, i.e., the product between the road lengths and capacities \cite{lonardi2021designing}.

Remarkably, adaptation equations allow extracting optimal fluxes efficiently \cite{lonardi2023immiscible,facca2021fast}, thus enabling good scalability to large networks and many OD pairs. In our algorithmic implementation, Kirchhoff's law is solved $M$ times, one per group of cyclists, with a sparse direct linear solver (UMFPACK, Unsymmetric MultiFrontal sparse LU Factorization \cite{davis2004algorithm}). Its complexity typically scales as $O(Z)$, where $Z$ is the number of non-zero entries of the network Laplacian $L^i$, and greatly benefits from the sparse character of $L^i$. Additionally, \Cref{eq: dynamics} has to be solved until a stopping criterion is met, for our case study, when the largest difference quotient between two consecutive $\mu$ and $J$ is below $\epsilon_\mu = \epsilon_J = 0.1$.

We test our algorithm on the experimental setup of Copenhagen's transportation network described in Secion 2.2 (main text) and compare it against a multi-source multi-sink Dijkstra's algorithm implementation where a single-source single-sink Dijkstra is repeated for all OD pairs. In Supplementary Information, we plot the total travel length $J(w=\ell)$ and the total travel cost $J(w)$ for the Optimal Transport dynamics and for Dijkstra's algorithm. Our method shows excellent agreement with Dijkstra. We also observe that the Optimal Transport adaptation dynamics retrieves the shortest paths in $t \sim 10$ min, whereas Dijkstra's algorithm takes $t \sim 1$ hour.

\subsection{Definitions of metric for bikeability}
We provide here the mathematical definitions of the metrics used in the main to evaluate network bikeability.

\paragraph{Overlap.}
The Overlap is defined as:
\begin{align}
    \text{Overlap}(\alpha) = \sum_{e \in E_2}\sum_{i = 1}^M \ell_e |F_e^i(w(\alpha))| \; \Big/ \sum_{e \in E_0}\sum_{i = 1}^M \ell_e |F_e^i(w(\alpha))| \,,
\end{align}
where $E_2$ is the set of bicycle edges, $E_0$ is the overall set of edges, and we made explicit all dependence on $w$ and, in turn, on $\alpha$.

When calculating the Overlaps for each network node, we first use: 
\begin{align}
    \text{Overlap}^i(\alpha) = \sum_{e \in E_2} \ell_e |F_e^i(w(\alpha))| \; \Big/ \sum_{e \in E_0} \ell_e |F_e^i(w(\alpha))| \quad \forall i = 1,\dots,M\,,
\end{align}
and then assign $\text{Overlap}^i(\alpha)$ to the origin nodes $O^i$, one for each group of cyclists. The Overlaps for all nodes $v$ that are not origin nodes are computed as weighted averages of $\text{Overlap}^i(\alpha)$:
\begin{align}
    \label{eq: overlap v}
    \text{Overlap}^v(\alpha) = \sum_i K(\text{dist}(v,i))\,  \text{Overlap}^i(\alpha) \; \Big/ \sum_i K(\text{dist}(v,i))  \quad \forall v \notin \{ O^i \}^i \,,
\end{align}
where $\text{dist}(v,i)$ is the Euclidean distance between $v$ and the origin node $O^i$ and $K$ is a weight function that we choose as an exponential $K(x) = \exp(- r x / \max_{u,v} \text{dist}(u,v))$, with a dimensionless coefficient set to $r = 150$.

\paragraph{Detour.}
The Detour is defined as:
\begin{align}
    \label{eq: detour}
    \text{Detour}(\alpha) = \left( \sum_{e \in E_0}\sum_{i = 1}^M \ell_e |F_e^i(w(\alpha))| \; \Big/ \sum_{e \in E_0}\sum_{i = 1}^M \ell_e |F_e^i(w = \ell)| \right) - 1\,.
\end{align}
Notice that, since the denominator inside the brackets in \Cref{eq: detour} computed for the optimal fluxes $F^\star$ is exactly the shortest length traveled by cyclists, then $\text{Detour}(\alpha) \geq 0$, and equality is met at $\alpha = 1$.

Similarly to the Overlap, at the node level, we compute:
\begin{align}
    \label{eq: detour node}
    \text{Detour}^i(\alpha) = \left( \sum_{e \in E_0} \ell_e |F_e^i(w(\alpha))| \; \Big/ \sum_{e \in E_0} \ell_e |F_e^i(w = \ell)| \right) - 1 \quad \forall i = 1,\dots,M
\end{align}
and average \Cref{eq: detour node} as in \Cref{eq: overlap v} for all nodes that are not origins.

\paragraph{Gini coefficient of fluxes.}

The Gini coefficient of an $N$-dimensional vector $x$ is defined as:
\begin{align}
    \text{Gini}(x) = \sum_{m,n=1}^N |x_m - x_n| \Big/ 2 N^2\Bar{x} \,,
\end{align}
where $\Bar{x}=\sum_{n=1}^N\, x_n/N$. In the main text, $x$ is the vector of total fluxes traveling on bicycle tracks and on roads. Hence, $x_n = \sum_i |F_e^i|$ and $N = |E_k|$, with $k=1,2$ being indexes denoting roads and bicycle tracks, respectively.

\section{Additional network visualizations}

\begin{center}
    \begin{figure}[H]
        \centering
        \includegraphics[width=1\textwidth]{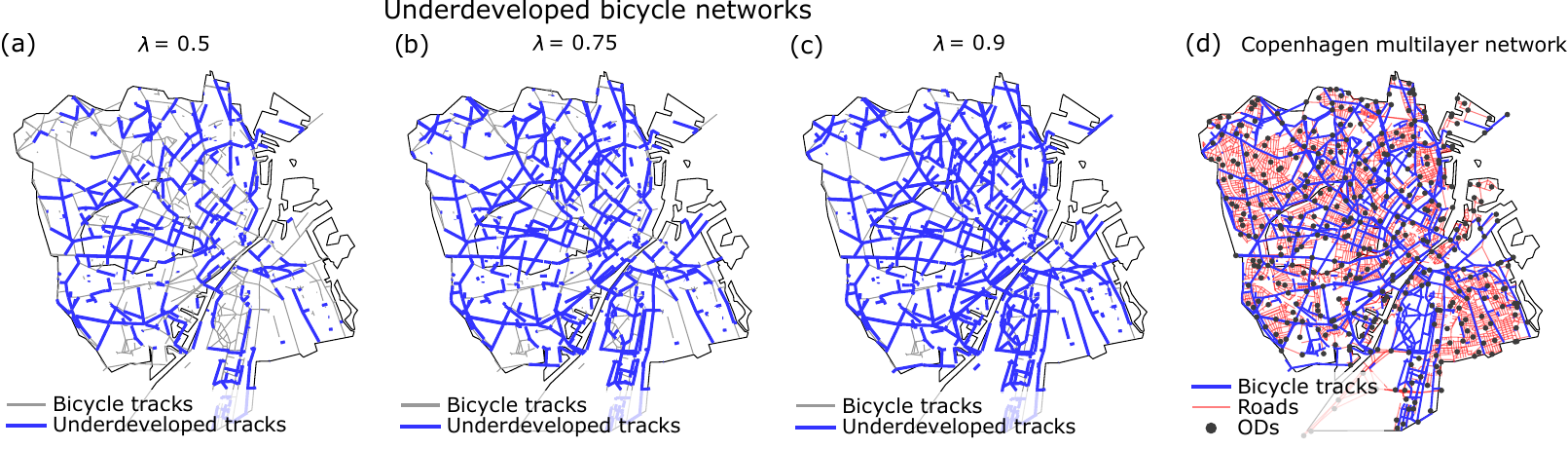}
        \caption{\textbf{Copenhagen synthetic network visualizations at different development stages.} Colors of nodes and edges are as Fig. 1(b) for (a)-(c) and as Fig. 1(a) for (d). \textbf{(a)} Unverdeveloped Copenhagen network for $\lambda = 0.5$, as in Fig. 1(b). \textbf{(b)} $\lambda = 0.75$. \textbf{(c)} $\lambda = 0.9$. \textbf{(c)} Copenhagen full network as in Fig. 1(a).}
        \label{fig: topologies supp}
    \end{figure}
\end{center}

In \Cref{fig: topologies supp}, we show underdeveloped Copenhagen's network topologies for different values of $\lambda$. These visualizations complement Fig. 1(a)-(b) (main text). We also provide additional network fluxes visualizations complementing those in Fig. 1(c)-(d) (main text). The plots show the fluxes on one random network for a fixed pair of  $(\lambda,\alpha)$, which spans $\lambda =  0.5, 0.75, 0.9, 1 $ and $\alpha = 0.5, 0.75, 0.95, 1$. In accordance with the results presented in the manuscript, we observe that lower values of $\alpha$ yield a greater concentration of fluxes on bicycle tracks. For $\lambda = 0.75,0.9$, we observe a high concentration of fluxes on the bicycle tracks. If the bicycle networks get too scarce, i.e., $\lambda = 0.5$, cyclists tend to travel on roads rather than protected tracks. This behavior underscores a crucial safety problem in cities like Zürich, shown in \Cref{fig: flux panel 5 supp}, where the bicycle tracks are largely underdeveloped (approximately 7\% of those of Copenhagen, see Table 1 (main text)). This means that even passengers with large incentives to ride on protected bicycle tracks ($\alpha = 0.5$) travel on roads when they do not have sufficient infrastructure at their disposal.

\section{Additional global metrics}

In \Cref{fig: cost panel supp}, we plot the total traveled length $J(w=\ell)$, and the total travel cost $J(w)$, with $w=\ell$ on roads and $w=\alpha \ell$ on bicycle tracks, for the whole network infrastructure made of cycling tracks and roads together. We compare (in \Cref{fig: cost panel supp}(a)) our Optimal Transport dynamics with Dijkstra's algorithm and find excellent agreement in the optimal values of the objective function $J(w)$ attained by the two methods. The main difference is the computational cost, with our OT-based algorithm taking a fraction of the time required to run Dijkstra, $t\sim 10min$ for our method against $t \sim 1 hour$, for Dijkstra's algorithm. In \Cref{fig: cost panel supp}(b), we observe that the total traveled length grows when $\alpha$ decreases. This is in accordance with the fact that cyclists detour onto longer paths as they prioritize safety (lower $J(w)$). As noted in the main manuscript, the total cycled length is not monotone with respect to $\lambda$. First, it increases from the full network to $\lambda = 0.9$, then decreases for lower values of $\lambda$.

\begin{center}
    \begin{figure}[H]
        \centering
        \includegraphics[width=1\textwidth]{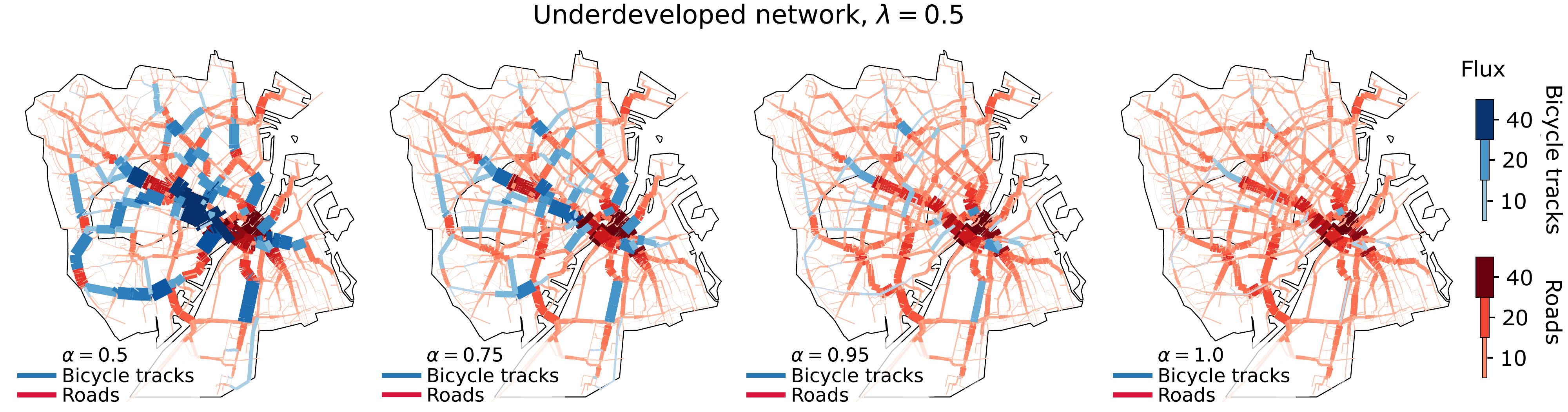}
        \caption{\textbf{Fluxes on the underdeveloped Copenhagen network for $\lambda = 0.5$.} Colors are as in Fig. 1(c)-(d) (main text).}
        \label{fig: flux panel 1 supp}
    \end{figure}
\end{center}

\begin{center}
    \begin{figure}[H]
        \centering
        \includegraphics[width=1\textwidth]{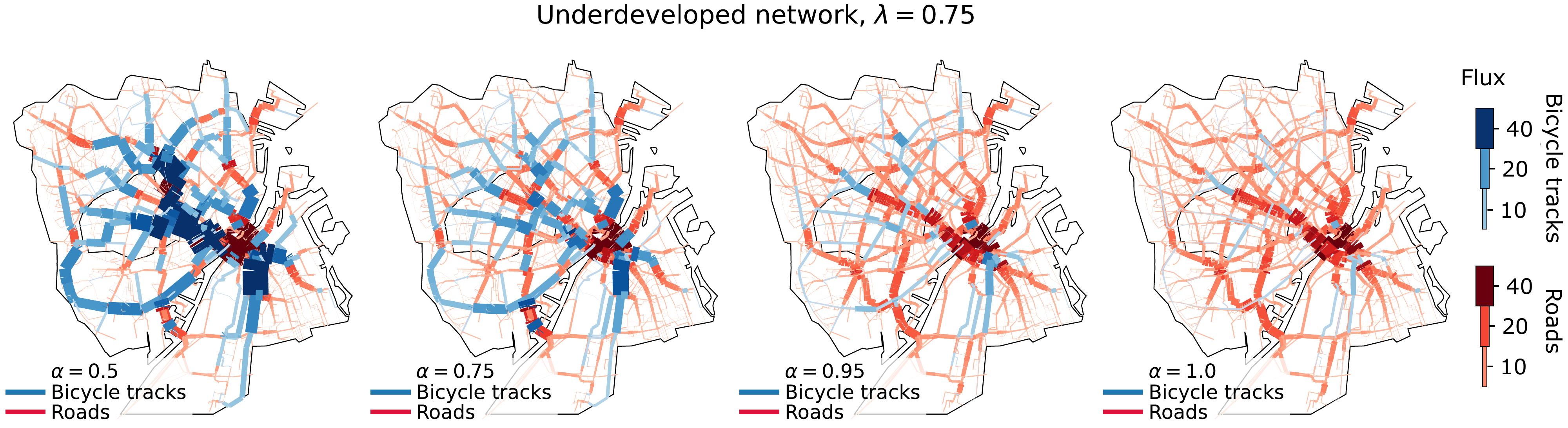}
        \caption{\textbf{Fluxes on the underdeveloped Copenhagen network for $\lambda = 0.75$.} Colors are as in Fig. 1(c)-(d) (main text).}
        \label{fig: flux panel 2 supp}
    \end{figure}
\end{center}

\begin{center}
    \begin{figure}[H]
        \centering
        \includegraphics[width=1\textwidth]{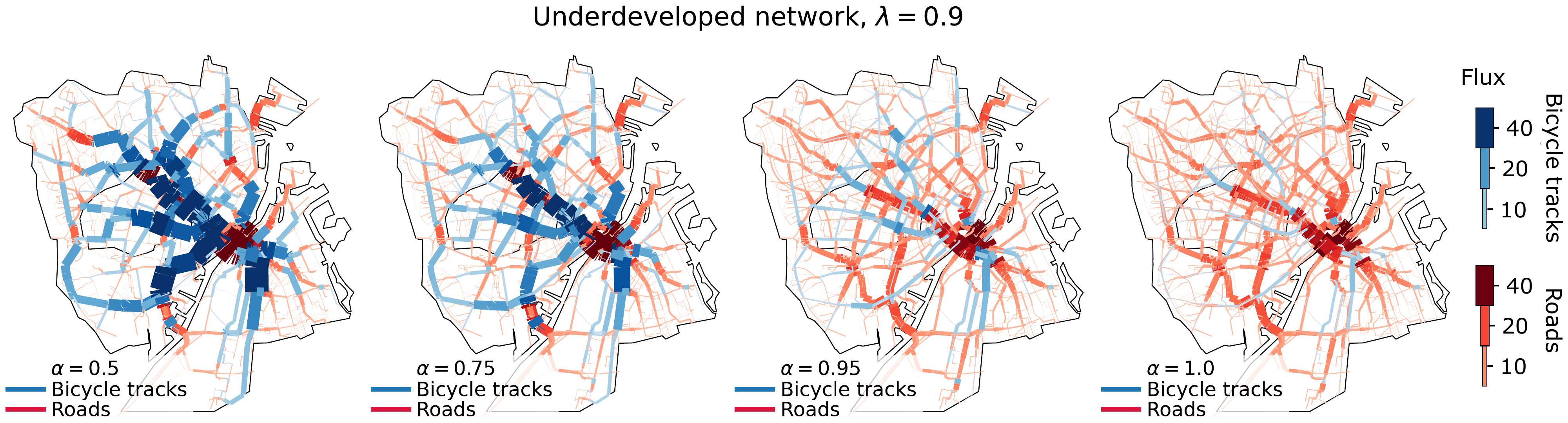}
        \caption{\textbf{Fluxes on the underdeveloped Copenhagen network for $\lambda = 0.9$.} Colors are as in Fig. 1(c)-(d) (main text).}
        \label{fig: flux panel 3 supp}
    \end{figure}
\end{center}

\begin{center}
    \begin{figure}[H]
        \centering
        \includegraphics[width=1\textwidth]{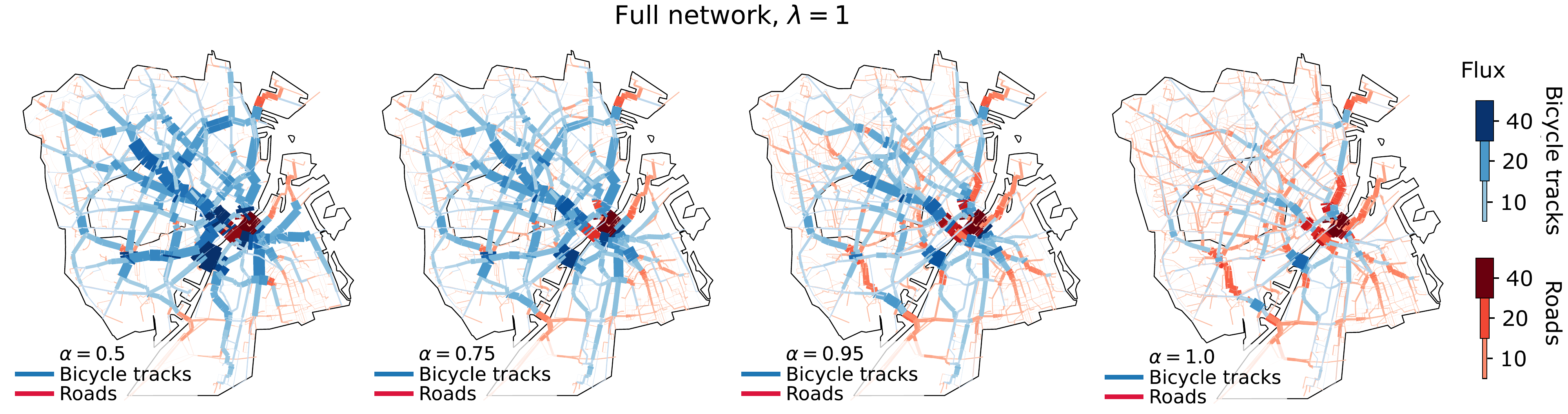}
        \caption{\textbf{Fluxes on the full network of Copenhagen for $\lambda = 1$.} Colors are as in Fig. 1(c)-(d) (main text).}
        \label{fig: flux panel 4 supp}
    \end{figure}
\end{center}

\begin{center}
    \begin{figure}[H]
        \centering
        \includegraphics[width=1\textwidth]{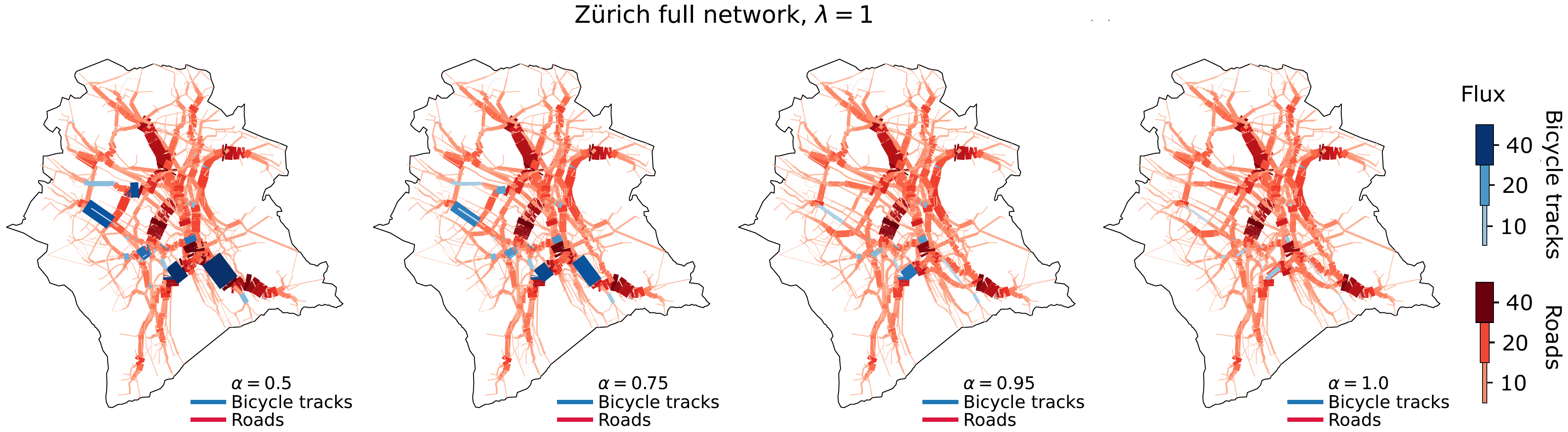}
        \caption{\textbf{Fluxes on the full network of Zürich.} Colors are as in Fig. 1(c)-(d) (main text).}
        \label{fig: flux panel 5 supp}
    \end{figure}
\end{center}

\begin{center}
    \begin{figure}[H]
        \centering
        \includegraphics[width=0.7\textwidth]{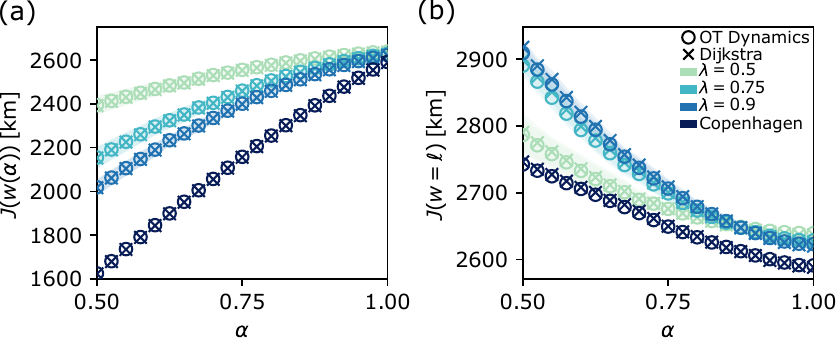}
        \caption{\textbf{Weighted travel cost $J(w(\alpha))$ and total traveled length $J(w=\ell)$ for our method and Dijkstra's algorithm.} Results for the Optimal Transport dynamics are in circles, whereas those for Dijkstra's algorithm are denoted with crosses. \textbf{(a)} Total travel cost $J(w)$ with $w=\alpha \ell$ on bicycle lanes and $w=\ell$ on roads. \textbf{(b)} Total traveled length $J(w=\ell)$. Companion Figure to Fig. 2 (main text).}
        \label{fig: cost panel supp}
    \end{figure}
\end{center}

\end{document}